# Amplitude- and gas pressure-dependent nonlinear damping of high-Q oscillatory MEMS micro mirrors

Ulrike Nabholz, Wolfgang Heinzelmann, Jan E. Mehner and Peter Degenfeld-Schonburg

*Abstract*—Silicon-based micro-electromechanical systems (MEMS) can be fabricated using bulk and surface micromachining technology. A micro mirror designed as an oscillatory MEMS constitutes a prominent example. Typically, in order to minimize energy consumption, the micro mirror is designed to have high quality factors. In addition, a phase-locked loop guarantees resonant actuation despite the occurrence of frequency shifts. In these cases, the oscillation amplitude of the micro mirror is expected to scale linearly with the actuation input power. Here however, we report on an experimental observation which clearly shows an amplitude depletion that is not in accordance with any linear behaviour. As a consequence, the actuation forces needed to reach the desired oscillation amplitude are by multiples higher than expected. We are able to explain the experimental observations accurately by introducing a single degree-of-freedom model including an amplitude-dependent nonlinear damping term. Remarkably, we find that the nonlinear damping shows a clear gas pressure dependency. We investigate the concepts and compare our findings on two different micro mirror design layouts.

*Index Terms*—microelectromechanical devices, nonlinear oscillators, microactuators, modeling

## I. INTRODUCTION

IN many applications, oscillatory micro-electromechanical systems (MEMS) are driven by an external periodic force and operated with a fixed phase relation between drive and system response [1]. Thus, for many resonant MEMS, a linear relation between oscillation amplitude and driving force can be assumed with sufficient accuracy. Advances in micromachining technology have opened up possibilities in the field of optical MEMS [2], [3]. A micro mirror constitutes a prominent example [4], [5].

In order to fulfil the requirements pertaining to functionality and performance of the MEMS micro mirror, high deflection angles have to be reached [6]. Compared to average MEMS devices, a micro mirror has rather large dimensions in the millimetre range [7]. Thus, high deflection angles correspond to large oscillation amplitudes for which the assumption of linear behaviour no longer holds true. The Duffing oscillator provides a basic nonlinear model that is widely used for resonant MEMS but it only accounts for linear damping [8] and cannot be used to emulate measured nonlinear damping phenomena which entail amplitude depletion.

In many cases, nonlinear damping increases power consumption and operating costs. Consequently, it is highly desirable to understand and avoid the causes for such energy dissipation.

The physical processes that are responsible for nonlinear damping are still the subject of an ongoing field of study in MEMS, NEMS as well as macro-scale systems [9], even though the influence of the suggested phenomenological nonlinear damping term has been analysed extensively [9], [10], [11].

Here, we report on experimental observations that show pronounced nonlinear damping effects in a MEMS micro mirror which depend on gas pressure, oscillation amplitude and design geometry. The observed phenomena can be explained by a single degree-of-freedom (DOF) model which we solve by adapting existing averaging methods [12]. The parameters of our model are found by data extraction from ring-down measurements of the micro mirror [13]. We highlight the amplitude- and pressure-dependency of nonlinear damping and analyse the effect of changes in device geometry on damping. In the test setup used, the internal pressure was varied between normal pressure and low vacuum. In addition, the experiments were performed on two design layouts with different spring geometries.

## II. EXPERIMENTAL SETUP

### A. Oscillating MEMS Micro Mirror

A micro mirror can be described as a scanning system with a torsional degree of freedom utilizing mechanical structures [14]. In the case of a 2D scanning system, the mirror can oscillate in two directions. Only the mechanics of one resonant axis will be modelled here. The system has been designed to achieve a deflection angle of $8°$. In the following analysis, the deflection at the outer edges of the reflective structure that corresponds to this angle is termed $a_{max}$.

Typical micro mirror designs employ either torsional springs or bending springs that encompass the actuator. Fig. 1 shows two designs of a tilting micro mirror: Fig. 1a provides a schematic of design layout A with bending springs, whereas Fig. 1b shows design layout B with torsional springs. The light grey area represents the reflective structure itself and the dark grey areas the surrounding frame, whereas a simplified form of the springs and their connection points to the mirror is outlined in red. The dashed line denotes the resonant axis around which the deflection of the mirror occurs.

U. Nabholz and P. Degenfeld-Schonburg are with Robert Bosch GmbH, Corporate Research, 71272 Renningen, Germany (e-mail: ulrike.nabholz@de.bosch.com).

W. Heinzelmann is with Robert Bosch GmbH, Automotive Electronics, 72762 Reutlingen, Germany.

J. E. Mehner is with Chemnitz University of Technology, 09107 Chemnitz, Germany.

The two designs are almost identical in size and working principle and vary mostly in their spring geometries (see Fig. 1): Design layout A that was investigated above employs springs that encompass the mirror and actuate the structure by bending. Consequently, since the springs are drawn outward away from the mirror, curve back in and attach to the structure from two sides, masses are added to the system further away from the resonant axis. In contrast, design layout B employs torsional springs where the mass is concentrated along the resonant axis.

Working principle and detailed graphical representations of designs using bending springs can be found in [15] and [16]. Torsional spring designs are widely known and their working principle has been illustrated numerous times, among others by [14].

The linear mode frequencies of the two design layouts are slightly different, with $f_{0,A} = 3200$ Hz, $f_{0,B} = 2946$ Hz. In order to characterize the micro mirror, the amplitude response curves for different actuation forces are recorded. Fig. 2 shows modelled amplitude response curves for six driving forces as they are expected from a conventional system model of a micro mirror based on the Duffing oscillator (introduced in section III-A). Between different levels of the input force the step size is kept constant. In this case, the maximum amplitude of the mirror deflection is expected to increase linearly with the input force.

From individually measured response curves of the system, the points of maximum amplitude are extracted. This yields the red data points given in Fig. 3 for design layout A, where the input force is normalized to the maximum value for this layout. The amplitude values correspond to the tips of the exemplary amplitude response curves in Fig. 2. A comparison of these points with the linear curve (black) shows that the system behaviour deviates from the linear expectation. This leads to the need for an enhanced system model in order to accurately emulate and predict the behaviour of the micro mirror.

Typically, the air pressure inside the packaging is lower than the ambient pressure surrounding the device. The test setup that was used for the measurements can be modified to actuate the mirror at different pressures in order to analyse the pressure-dependency of the system response. For the variation of internal pressure, a micro mirror inside a pressure cell is set to an ambient pressure between 5 mbar and normal pressure.

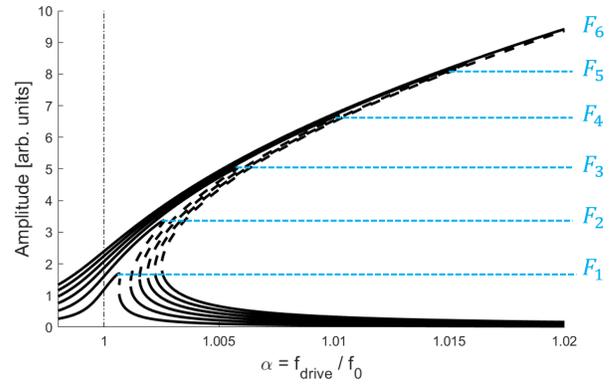

Fig. 2. Exemplary amplitude response curves for different driving forces ($F_1 - F_6$) for a conventional Duffing oscillator model. The horizontal axis shows the dimensionless detuning factor $\alpha$, the vertical axis shows the amplitude normalized to its maximum value. Stable solution branches denoted by solid lines, unstable solution branches denoted by dashed lines. A higher driving force increases the overall level of the amplitude response as well as the frequency shift that is reached at the peak of the curve. The relation between input force and maximum amplitude is linear in this model.

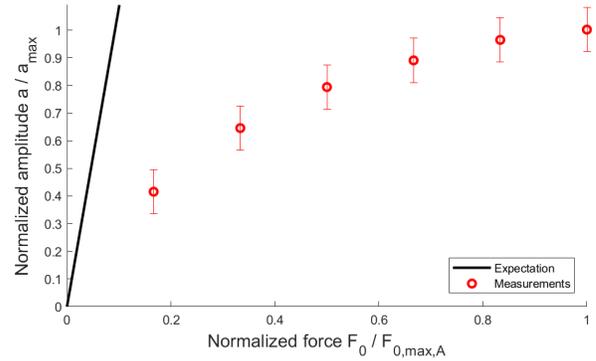

Fig. 3. Expected and measured values for the maximum points of the amplitude response curves for a wide range of input forces in design layout A. The system behaviour measured at several values of the input force (denoted by the red circles) clearly deviates from the expected linear behaviour (black). For higher forces, the measured amplitudes level off, indicating a deviation from the linear model.

### B. Ringdown Response

Ring-down behaviour can be measured as follows: A frequency generator is set to the linear frequency of the desired mode. The oscillation amplitude is measured as a function of the excitation frequency. The drive frequency is ramped up and thus the amplitude of oscillation increases. When the measured amplitude exceeds a threshold value, the external driving force is switched off and the amplitude decay is recorded until the oscillation has abated and only a noise signal remains. The signal is given by a voltage that is proportional to the deflection of the micro mirror which is measured using piezo-resistors connected to the mirror's suspension. The sampling is performed without any discernible delay and at a rate that is sufficient for logging the signal as a function of time. In the following, all oscillation amplitudes are normalized to the initial value $a_0$.

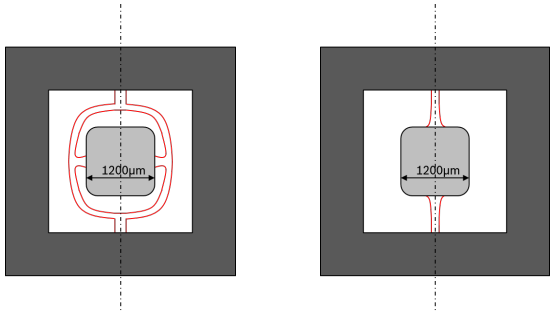

Fig. 1. Schematic representations of design layouts A and B: In both layouts, the light grey area represents the reflective mirror surface, the dark grey areas represent the frame surrounding the mirror. The springs are outlined in red and the dashed vertical line indicates the position of the resonant axis. Fig. 1a shows design layout A with bending springs that encompass the reflective structure and attach to it from the side. Fig. 1b shows design layout B with torsional springs that attach to the reflective structure in line with the resonant axis. Both design layouts are surrounded by a package of the same dimensions.

Fig. 4a shows the portion of the signal that is analysed starting from the point in time when the driving force is switched off. Fig. 4b depicts several exemplary oscillations.

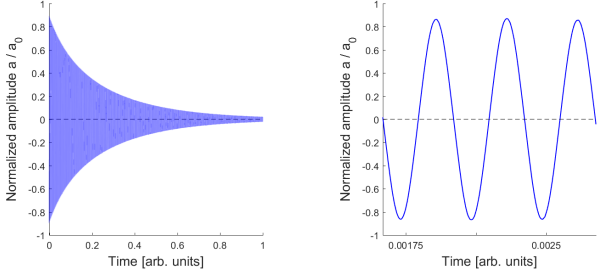

Fig. 4. Measured signal from ring-down response. Fig. 4a shows the amplitude of the full ring-down signal over time normalized to its initial value $a_0$. Due to the speed of oscillation, individual oscillations are not visible in this representation. Fig. 4b shows an excerpt of this data for only three full oscillations of the system. In both cases, the time scale of the oscillation is given in arbitrary units.

## III. METHODS

### A. Mechanical system model

The simplest equation of motion is given by the harmonic oscillator as

$$\ddot{q} + \omega_0^2 q + \frac{\omega_0}{Q}\dot{q} = F_0 \cdot \sin(\omega_d t) \quad (1)$$

In (1), $q = q(t)$ is the time-dependent modal coordinate of the oscillation. Consequently, $\dot{q} = \dot{q}(t)$ denotes the speed and $\ddot{q} = \ddot{q}(t)$ the acceleration. $F_0$ is the amplitude of the specific driving force. The angular frequencies $\omega_0$ and $\omega_d$ are the linear mode frequency and the driving frequency, respectively. The linear mode frequency, $\omega_0$, is a function of Young's Modulus, Poisson ratio and geometry. Thus, it is inherent in the mechanical system and independent of external influences. The frequencies $f_0$ and $f_d$ are linked to the angular frequencies with the same indices by the relation

$$\omega = 2\pi f. \quad (2)$$

The linear quality factor $Q$ is inversely proportional to the damping. Therefore, for an undamped system, it approaches infinity and $\frac{\omega_0}{Q} \to 0$. The quality factor can be depicted as the ratio of energy contained in the system to the energy that is dissipated in each cycle.

The three terms on the left hand-side of (1) can be categorised as follows: $\ddot{q}$ and $\omega_0^2 q$ are purely structural terms, whereas the term $\frac{\omega_0}{Q}\dot{q}$ adds linear damping.

From experience, it can be assumed that the harmonic oscillator model as given in (1) is not sufficient for the modelling of a micro mirror. A widely used model in this context, the Duffing oscillator [8], comprises an additional term, the cubic nonlinearity $\beta q^3$, where $\beta$ is the Duffing coefficient. This coefficient is dependent on the geometry of the system and thus used to model geometric nonlinearities.

$$\ddot{q} + \omega_0^2 q + \frac{\omega_0}{Q}\dot{q} + \beta q^3 = F_0 \cdot \sin(\omega_d t) \quad (3)$$

From the results of measurements as the ones shown in Fig. 3, it can be assumed that further terms are needed to accurately model the system behaviour of a micro mirror. In order to ascertain whether other damping effects apart from linear damping alter the frequency response, a phenomenological equation of motion is introduced:

$$\ddot{q} + \omega_0^2 q + \frac{\omega_0}{Q}\dot{q} + \beta q^3 + \frac{\omega_0}{Q_{nl}}q^2\dot{q} = F_0 \cdot \sin(\omega_d t). \quad (4)$$

It expands the basic Duffing oscillator given in (3) by the additional damping term $\frac{\omega_0}{Q_{nl}}q^2\dot{q}$. The quantity $Q_{nl}$ is termed the nonlinear quality factor and as such it is inversely proportional to nonlinear damping.

Since an analytical solution for the steady-state of the micro mirror system is desired for further analysis, the method of averaging [12] will be used. The system behaviour exhibits two time scales: The time scale $t$ depicts fast changes, whereas $\tau$ changes slowly. The two time scales are related by the linear mode frequency $\omega_0$:

$$t = \frac{\tau}{\omega_0}. \quad (5)$$

The detuning factors $\alpha = \frac{\omega_d}{\omega_0}$ and $\sigma = (1 - \alpha^2)$ are introduced into (4) resulting in

$$\frac{d^2q}{d\tau^2} + \alpha^2 q = -\sigma q - \frac{1}{Q}\frac{dq}{d\tau} - \frac{\beta}{\omega_0^2}q^3 - \frac{1}{Q_{nl}}q^2\frac{dq}{d\tau} + \frac{F_0}{\omega_0^2}\sin(\alpha\tau), \quad (6)$$

where $q = q(\tau)$. As the first step in the method of averaging, the method of variation of parameters is applied [17]: The dependent variable $q$ is split up into amplitude $a$ and phase relation between drive and system response $\phi$ of the oscillation. Independent of the equation of motion, using the slow time scale, the modal coordinate is given by

$$q(\tau) = a(\tau)\sin(\alpha\tau + \phi(\tau)) \quad (7)$$

and its time-derivation yields

$$\frac{dq(\tau)}{d\tau} = \alpha\, a(\tau)\cos(\alpha\tau + \phi(\tau)). \quad (8)$$

The following short notations for the time-dependent variables are introduced:

$$a := a(\tau) \text{ and } \dot{a} := \frac{da(\tau)}{d\tau}, \quad (9)$$

$$\phi := \phi(\tau) \text{ and } \dot{\phi} := \frac{d\phi(\tau)}{d\tau}, \quad (10)$$

$$q := q(\tau) \text{ and } \dot{q} := \frac{dq(\tau)}{d\tau}. \quad (11)$$

In order to simplify the notation, we introduce the substitution $\theta = \alpha\tau + \phi(\tau)$. The differential equations for amplitude $\dot{a}$ and phase $\dot{\phi}$ for the phenomenological equation of motion (4) can be formulated as [12]

$$\dot{a} = \frac{1}{2\pi\alpha} \int_0^{2\pi} f(q, \dot{q}) \cos(\theta) d\theta \tag{12}$$

$$= -\frac{a}{2Q} - \frac{F}{\omega_0^2} \frac{\sin(\phi)}{2\alpha} - \frac{a^3(\tau)}{8Q_{nl}}, \tag{13}$$

$$\dot{\phi} = \frac{1}{2\pi\alpha} \int_0^{2\pi} f(q, \dot{q}) \left(-\frac{1}{a}\sin(\theta)\right) d\theta \tag{14}$$

$$= \frac{\sigma}{2\alpha} - \frac{F}{2\alpha a \omega_0^2} \cos(\phi) + \frac{3\beta a^2}{8\alpha\omega_0^2}. \tag{15}$$

By definition, the steady-state is reached for

$$\dot{a} = \dot{\phi} = 0. \tag{16}$$

The steady-state amplitude can be determined by converting (13) into phase

$$\tan(\phi) = \frac{-\frac{1}{2Q} - \frac{a^2(\tau)}{8Q_{nl}}}{\frac{\sigma}{2\alpha} + \frac{3\beta a^2(\tau)}{8\alpha\omega_0^2}} \tag{17}$$

and amplitude

$$a^2 = \frac{\left(\frac{F}{2\alpha\omega_0^2}\right)^2}{\left(\frac{\sigma}{2\alpha} + \frac{3\beta a^2}{8\alpha\omega_0^2}\right)^2 + \left(\frac{1}{2Q} + \frac{a^2}{8Q_{nl}}\right)^2}. \tag{18}$$

Note that (18) is not an explicit function of the amplitude. Its numerator shows the direct influence of the driving force on the amplitude. The denominator is dimensionless and contains the interdependencies between amplitude and frequency: The left addend yields the frequency shift caused by the detuning between the driving frequency $\omega_d$ and the linear frequency $\omega_0$, as well as the Duffing shift.
The right addend contains the quality factors and thus influences the prominence of the resonance peak: Through the additional nonlinear damping term that depends on the oscillation amplitude, the nonlinear quality factor $Q_{nl}$ reduces the effective quality factor of the system.
In order to ensure the accuracy of the steady-state solutions obtained using the method of averaging, they are contrasted with the results of a transient numerical simulation of the equation of motion (4). Fig. 5 shows the stable steady-state solutions in black and the unstable solution in red (stability properties have been obtained from a stability analysis of the system [18]), as well as the results of the transient simulation shaded in red. The transient model contains the equation of motion and a sinusoidal input force with a variable frequency: At the initial frequency, the system starts out on the upper stable branch in Fig. 5. When the frequency is ramped up, it eventually reaches the tip of the response curve, where this solution branch becomes unstable and the system response drops down onto the lower stable branch. Thus, the steady-state solution obtained using the method of averaging is in very good agreement with the numerical simulation of the full equation of motion.
For an infinitely slow frequency ramp, the envelope data of the transient simulation would be equivalent to the steady-state solutions. Due to dynamic effects [19], all realistic frequency ramps will show some deviation from the steady-state values.

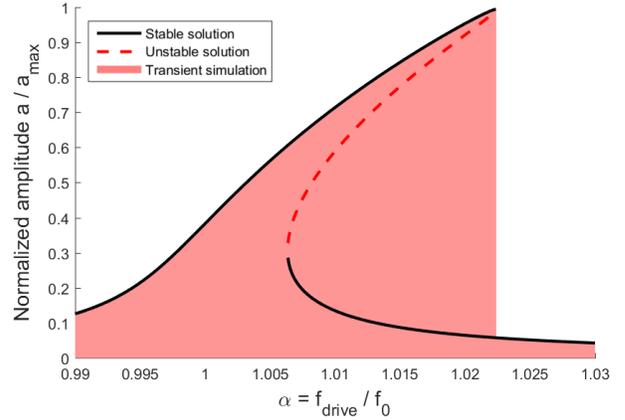

Fig. 5. Model validation using a transient simulation of the full system given in (4): The frequency of the sinusoidal input force is ramped up slowly. The solid black lines denote the stable steady-state solutions, the dashed red line the unstable steady-state solution and the shaded area shows the full oscillation of the transient system response. Evidently, the steady-state solution obtained from the method of averaging is in excellent agreement with the transient solution.

*B. Data extraction procedure*

We follow the lines of [13] to extract the linear quality factor $Q$, nonlinear quality factor $Q_{nl}$, linear mode frequency $\omega_0$ and Duffing coefficient $\beta$ from ring-down measurements. Since the external signal is switched off prior to the start of the recording of the ring-down signal, the equation of motion given by (4) is simplified to

$$\ddot{q} + \omega_0^2 q + \frac{\omega_0}{Q}\dot{q} + \beta q^3 + \frac{\omega_0}{Q_{nl}} q^2 \dot{q} = 0 \tag{19}$$

The basis of the subsequent analysis is provided by the general solution to the equation of motion in (19) given by

$$q = q(t) = a(t) \sin(\omega_0 t + \phi(t)) \tag{20}$$

For the initial condition of the phase relation $\phi(t)$, we set $\phi(t=0) = 0$. Thus, the solution provided by the method of averaging is obtained by simplifying (13) and (15) for $F_0 = 0$. This yields

$$\dot{a}(\tau) = -\left(\frac{1}{2Q} + \frac{a^2}{8Q_{nl}}\right) a, \tag{21}$$

$$\dot{\phi}(\tau) = \frac{3\beta a^2}{8\omega_0^2}. \tag{22}$$

The time-dependent solutions are known [13] and given by

$$a(t) = \frac{a_0 e^{-\frac{\omega_0}{2Q}t}}{\sqrt{g(t)}}, \tag{23}$$

$$\phi(t) = \frac{3\beta Q_{nl}}{2\omega_0^2} \ln(g(t)), \tag{24}$$

with

$$g(t) = 1 + \frac{Q}{4Q_{nl}} a_0^2 \left(1 - e^{-\frac{\omega_0}{Q}t}\right) \tag{25}$$

We define a long-time limit (denoted by the index "ltl"): For $\frac{\omega_0}{2Q} t \gg 1$, $e^{-\frac{\omega_0}{2Q} t}$ is rendered small and linear amplitude decay can be assumed. Therefore

$$g(t) = 1 + \frac{Q}{4Q_{nl}} a_0^2 \left(1 - e^{-\frac{\omega_0}{Q}t}\right) \approx 1 + \frac{Q}{4Q_{nl}} a_0^2 =: g_{ltl} \quad (26)$$

Thus, the amplitude in the long-time limit simplifies to

$$a_{ltl}(t) = \frac{a_0 \, e^{-\frac{\omega_0}{2Q}t}}{\sqrt{g_{ltl}}}. \quad (27)$$

The assumptions for the long-time limit and the simplifications they entail in (27) will be used for fitting the linear and nonlinear quality factors.

*C. Parameter Fits*

The envelope data of the ring-down signal shown in Fig. 4a corresponds to the deflection angles and hence the amplitudes for each cycle. Due to the presence of noise that interferes with the evaluation at low amplitudes, a threshold value is defined as $a_{thr} = \frac{1}{10} a_0$. Going backwards in time from the threshold time for a given amount of cycles, a section of linear amplitude decay is utilized for obtaining the linear quality factor $Q$. Taking the logarithm of (27) yields

$$ln\left(\frac{a_{ltl}(t)}{a_0}\right) = -\frac{\omega_0}{2Q} t - ln\left(\sqrt{g_{ltl}}\right). \quad (28)$$

The linear quality factor $Q$ is related to the slope of the curve, whereas the nonlinear quality factor $Q_{nl}$ is proportional to the y-axis intercept. Using the approximations for $Q$ and $Q_{nl}$, the time-dependent amplitude can be calculated when $\omega_0$ is known from the phase information obtained below. The results of the fit are shown in Fig. 6. The fractions $\frac{\omega_0}{Q}$ and $\frac{\omega_0}{Q_{nl}}$ correspond to linear and nonlinear damping of the system, respectively and are known from the analysis above. Thus, all coefficients in (25) and (23) are defined.

becomes clear that the phenomenological equation (4) yields a much better approximation of the ring-down curve. Fig. 6 displays a logarithmic scale in order to better illustrate the deviation from the exponential decay expected from a Duffing oscillator (grey line). For small amplitudes, the decay becomes close to exponential - it continues parallel to the Duffing fit. For the early phase of the decay the curve clearly differs from the exponential form. Fig. 6 shows data and fit functions for design layout A at an internal pressure of 125 mbar. The results for different internal pressures correspond qualitatively to the ones obtained for this exemplary pressure. Thus, we clearly see the presence of nonlinear amplitude-dependent damping in the MEMS micro mirror.

Obtaining the Duffing coefficient $\beta$ requires extraction of the phase information from the signal data in order to approximate the frequency as a function of time. The frequency for each cycle is determined from the zero-crossings of the signal data as pictured in Fig. 4b. The discrete nature of the signal leads to a discretisation error that is ameliorated by fitting a sine function to the oscillation period and extracting the frequency. For smaller amplitudes, the frequency is inaccurately represented, since the signal becomes more distorted by background noise. As a remedy, the frequency is averaged over several oscillation periods, thus smoothing out the data in order to attain high accuracy and viable precision.

The time-dependent frequency $\omega(t)$ can be written as

$$\omega(t) = \omega_0 + \dot{\phi}, \quad (29)$$

where the phase shift $\dot{\phi}$ is given by (22). The linear mode frequency $\omega_0$ is equal to the frequency in the long-time limit. Consequently, (22) can be used to fit the Duffing coefficient $\beta$. Fig. 7 shows the normalized frequency for design layout A

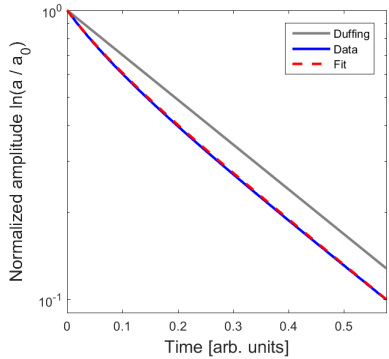

Fig. 6. Normalized amplitude data (blue) compared to fit function (dashed red) for design layout A at 125 mbar internal pressure. The basic Duffing oscillator with linear damping (grey) is displayed for comparison. The horizontal axis shows the time scale in arbitrary units, the vertical axis shows the logarithmic oscillation amplitude of the system normalized to its initial value $a_0$.

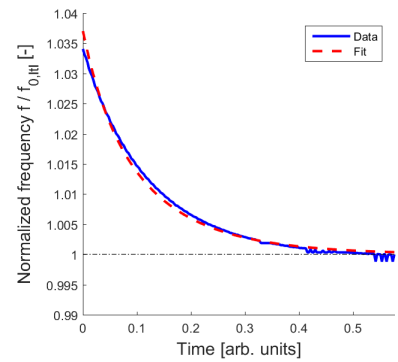

Fig. 7. Normalized frequency data (blue) compared to fit function (dashed, red) for design layout A at 125 mbar internal pressure. The horizontal axis shows the time scale in the same arbitrary units introduced in Fig. 4. The vertical axis shows the oscillation frequency of the system normalized to the value it approaches in the long-time limit, $f_{0,ltl}$.

The blue line in Fig. 6 denotes the data obtained from the ring-down measurements, whereas the dashed red line shows the fit function. The grey line provides the fit function for a basic Duffing oscillator with linear damping for comparison. It at an internal pressure of 125 mbar as a function of time. The blue line denotes the measured data averaged over several oscillation periods, the dashed red line shows the fit function given by (29).

## IV. RESULTS

### A. Pressure Dependency of Nonlinear Damping

The ring-down measurements were carried out and analysed for seven internal pressures ranging from normal pressure at 970 mbar down to 5 mbar for design layout A and B. Extracting the coefficients $Q, Q_{nl}, \omega_0$ and $\beta$ for each data set leads to an equation of motion in the form of (4) for each pressure with fully determined coefficients.

Fig. 8 gives an overview of the empirical relations found for different pressures in design layout A and B. Both coefficients are plotted against the logarithmic pressure and normalized to the values extracted for an internal pressure equal to normal pressure.

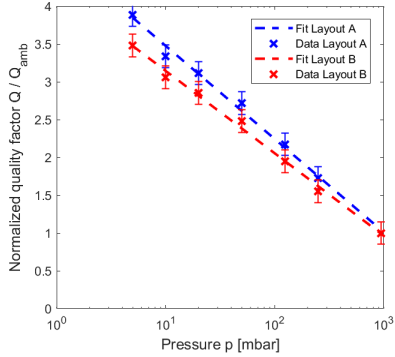

(a) Linear quality factors

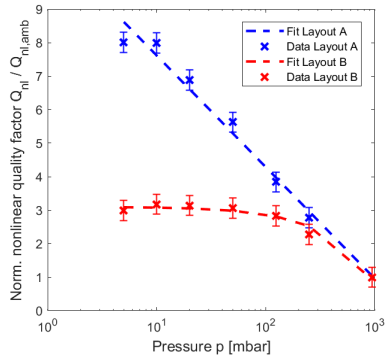

(b) Nonlinear quality factors

Fig. 8. Empirical relations between coefficients and pressure (data points and fit functions for design layout A in blue, for design layout B in red). The horizontal axis in each subfigure shows the ambient pressure in mbar on a logarithmic scale. The pressure ranges from 5 mbar to normal pressure at 970 mbar. All values on the horizontal axis have been normalized to their respective values at ambient pressure (Layout A: $Q_{amb} = 720, Q_{nl,amb} = 4.87$, Layout B: $Q_{amb} = 1120, Q_{nl,amb} = 76.3$). Fig. 8a shows the linear quality factor $Q$ on a logarithmic scale, Fig. 8b shows the nonlinear quality factor $Q_{nl}$ on a logarithmic scale. The fit functions used for the linear and nonlinear quality factors of both designs serve as a guide to the eye.

Figures 8a and 8b show the quality factor $Q$ and the nonlinear quality factor $Q_{nl}$, respectively. Crosses denote the values extracted from the measurements, whereas the dashed fit functions serve as a guide to the eye.

When comparing the coefficients for the two layouts, we note that the functional dependency of the nonlinear quality factor of design layout B deviates strongly from layout A. While the nonlinear damping of layout A seems to be dominated by gas pressure, the nonlinear damping of design layout B for pressures below 100 mbar is more likely governed by nonlinear friction [13], [20].

The Duffing coefficient $\beta$ and the linear mode frequency, $f_0$ remain constant over the whole pressure range. This behaviour is to be expected, since both the Duffing coefficient and the linear mode frequency of an oscillatory system are solely geometry-dependent.

The absolute values of linear quality factor $Q$ as well as nonlinear quality factor $Q_{nl}$ are higher for design layout B, corresponding to a lower amount of damping. The different amounts of linear damping become apparent by contrasting the amplitude response curves of both design layouts with linear and nonlinear damping in Fig. 9 in anticipation of the results in the next section: A higher quality factor and thus lower damping is equivalent to a more narrow amplitude response curve. Likewise, comparing design layout A and B for PLL operation in Fig. 11 shows that design layout A exhibits stronger nonlinear damping as well.

The Duffing coefficient $\beta$ stays constant for different pressures as was the case in design layout A, but design layout B shows a much smaller Duffing nonlinearity.

### B. Influence of Nonlinear Damping on System Response

Through the analysis of the ring-down measurements, the coefficients in the equation of motion (4) are fully defined and the influence of nonlinear damping can thus be quantified.

In Fig. 9, the amplitude response curves for linear and nonlinear damping are compared for both design layouts. Without the inclusion of nonlinear damping into the equation of motion, the simulation of a system with realistic parameters yields a detuning factor at resonance, $\alpha_{res}$, that is far higher than observed in measurements, especially in design layout A. Fig. 9a shows how the maximum amplitude and maximum detuning factor diminish, when nonlinear damping is introduced into the system. The effect of nonlinear damping on the amplitude response curve is much less pronounced in design layout B as shown in Fig. 9b.

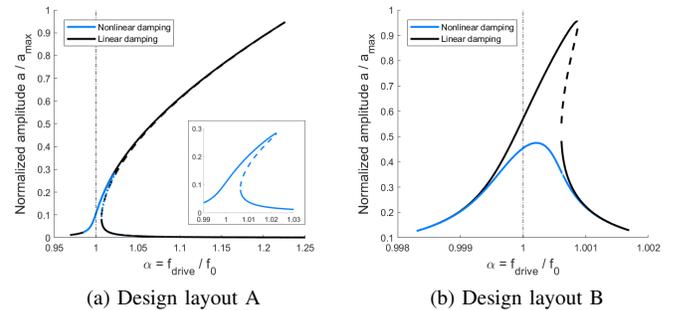

(a) Design layout A    (b) Design layout B

Fig. 9. Comparison of the amplitude response curves for linear damping (black) and nonlinear damping (blue) in design layouts A and B. The solid lines denote stable steady-state solutions, whereas the dashed lines denote unstable ones. The horizontal axes show the dimensionless detuning factor $\alpha$. For a value of $\alpha = 1$, the drive frequency is equal to the linear mode frequency. In Fig. 9a for design layout A, the addition of nonlinear damping to the system significantly reduces the maximum amplitude and the maximum frequency shift that is reached. The inset shows a more detailed view of the system with nonlinear damping. In Fig. 9b, nonlinear damping also reduces the maximum amplitude and the maximum frequency shift that is reached. It counteracts the Duffing nonlinearity and thus, the bistable region vanishes in the case of nonlinear damping. Compared to Fig. 9a for design layout A, the influence of nonlinear damping on the system behaviour is much smaller.

The influence of nonlinear damping on the excitation force needed to achieve a specific amplitude, for a fixed detuning factor $\alpha = 1.016$ is shown in Fig. 10a for design layout A. We find that the famous S-shape curve, found in many other fields of science [21], [22], [23], [24], is not altered significantly in the presence of nonlinear damping.

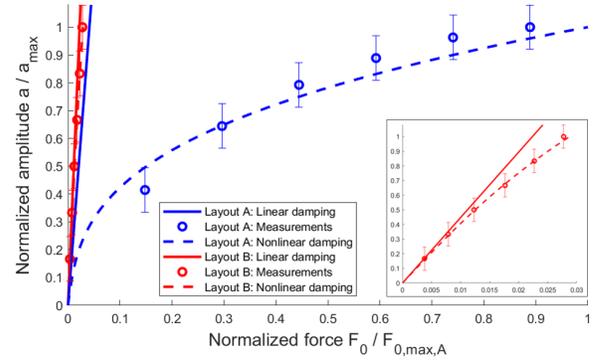

Fig. 11. Amplitude of the system using a PLL with nonlinear damping (dashed lines) and with linear damping (solid lines) at normal pressure (970 mbar) for design layout A (blue) and B (red). Individual points have been measured for comparison. The horizontal axis shows the input force normalized to its maximum values for design layout A. The vertical axis shows the amplitude normalized to the required deflection angle of the micro mirror. The maximum amplitude is equal for both design layouts, whereas the maximum input force for design layout A, $F_{0,max,A}$ is much higher than for design layout B. The inset shows a detailed view of the range of input forces relevant to design layout A. For design layout A, the amplitude of the nonlinearly damped system levels off at higher input forces. The nonlinear damping model is in good agreement with the measurements. For design layout B, the effect of nonlinear damping is less pronounced than for design layout A. For lower ambient pressures, the modelled curves show the same qualitative behaviour as can be deduced from Fig. 8.

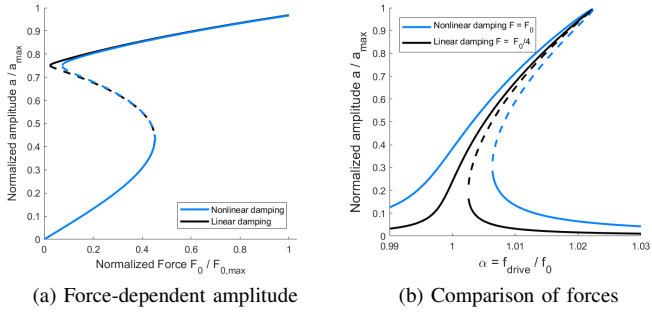

(a) Force-dependent amplitude  (b) Comparison of forces

Fig. 10. Influence of the driving force for linear damping (black) and nonlinear damping (blue) in design layout A for $\alpha = 1.016$. In Fig. 10a, the horizontal axis shows the input force normalized to its maximum value, the vertical axis shows the steady-state amplitude also normalized to its maximum value. The solid lines denote stable solutions whereas the dashed line denotes the unstable solution branch. In this representation, the difference between the linearly and nonlinearly damped system appears small. The influence of nonlinear damping on the system response, however, is more prominent in Fig. 10b. The horizontal axis shows the detuning factor $\alpha$, the vertical axis shows the steady-state amplitude normalized to its maximum value. In the case of nonlinear damping, shown in blue, an input force $F_0$ is chosen. For linear damping, shown in black, the input force is chosen in such a way that the maximum amplitude reached is approximately the same as in the nonlinear case. For this, the input force necessary lies at roughly $\frac{F_0}{4}$. This representation emphasizes the effect that nonlinear damping has on the necessary input force for a certain amplitude.

Fig. 10b shows that in order to reach the same maximum amplitude in the given system with nonlinear damping, a fourfold increase of the driving force is necessary. The factor of increase in the driving force depends on the individual system and the target amplitude. Thus, nonlinear damping decreases the effective quality factor of an oscillator, also apparent from the broadening of the response curve in Fig. 10b.

### C. Nonlinear Damping Under Realistic Operating Conditions

Under realistic operating conditions, stable resonant actuation of the micro mirror is ensured by using a phase-locked loop (PLL) [25] that keeps the system response at the very tip of the amplitude response curve for each given input force as shown in Fig. 2 and 3.

Fig. 11 shows the calculated phase-locked system for both nonlinear and linear damping (as already introduced in Fig. 3) compared to individually measured values for both design layouts.

Comparing the force that is necessary to achieve the same deflection angle in a system with linear as opposed to nonlinear damping showcases the adverse effect of nonlinear damping: In order to reach the same deflection angle, a much higher force, and thus, voltage, has to be applied.

For PLL operation of design layout B, also shown in Fig. 11, there is no significant difference in the frequency shift between the basic Duffing oscillator model and the nonlinear damping model. The dashed curve that denotes nonlinear damping levels out slightly at higher deflection angles, but compared to design layout A, the influence is negligible and the PLL behaviour can be assumed as linear in good approximation.

In both design layouts, the maximum amplitude used for normalization, $a_{max}$, is identical and corresponds to the target amplitude of the micro mirror system. The necessary input force is different for the two design layouts and the comparison is made using the maximum input force for design layout A, $F_{0,max,A}$, as a reference point.

Consequently, energy dissipation induced by amplitude- and pressure-dependent nonlinear damping can play an important role in the operation of torsional micro mirrors and depends on the geometry of the design layout.

## V. CONCLUSION

We reported on experimental observations highlighting the presence of amplitude- and gas pressure-dependent nonlinear damping in high-Q oscillatory MEMS micro mirrors. Most significantly, experiments show an oscillation amplitude depletion of the micro mirror which is not in accordance with the expectation of a linear dependence between oscillation amplitude and actuation force. As a consequence, the actuation forces required to reach a certain target oscillation amplitude need to be by several multiples higher than expected.

We were able to model all the observed phenomena by introducing a single degree-of-freedom Duffing oscillator model with an additional nonlinear damping term that comprises a nonlinear quality factor. From the equations of motion of our micro mirror model we derived algebraic steady-state expressions that show a decrease of the effective quality factor of the system caused by the nonlinear damping term.

The parameters of the nonlinear system model were fitted using data from ring-down measurements for ambient gas pressure ranging from 5mbar to normal pressure. As expected, the Duffing coefficients and linear mode frequency are gas pressure-independent whereas the quality factor depends on the pressure inside the micro mirror package. Remarkably, we found that the nonlinear quality factor also depends on pressure from which we conclude that nonlinear gas damping plays an important role and in most of our observations even dominates over nonlinear friction.

In addition, we have investigated our concepts on two different micro mirror spring designs where we found distinct quantitative results with one design being much more affected by nonlinear damping.

Since nonlinear damping phenomena as the one here considered can lead to much higher power consumption and operating costs it is highly desirable to investigate the physical origins of these effects. In this work, we exemplified that gas-solid interactions can be responsible for amplitude-dependent nonlinear damping and even dominate over nonlinear friction. One possible origin of nonlinear damping is given by squeeze film damping [26] that can occur in micro mirrors [27]. Thus, fundamental research on gas-solid interactions could enable theoretical predictions of nonlinear damping for arbitrary geometry design.

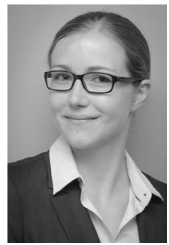
**Ulrike Nabholz** received the B.S. and M.S. degree in mechanical engineering from the Karlsruhe Institute of Technology, Karlsruhe, Germany, in 2014 and 2017, respectively.

She is currently conducting research with the Department of Microsystems and Nanotechnology at Robert Bosch GmbH, Corporate Research, Renningen, Germany, whilst pursuing a Ph.D. degree with the Department of Microsystems and Semiconductor Technology at Chemnitz University of Technology, Chemnitz, Germany.

Her research is focused on nonlinear dynamics in resonant MEMS and the development of reduced order models.

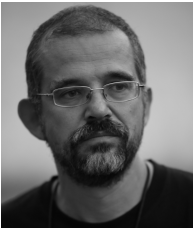

**Wolfgang Heinzelmann** received the diploma in electrical engineering from the University of Karlsruhe, Karlsruhe, Germany in 1994.

He is currently a pre-development engineer with the Department of Engineering Advanced MEMS Concepts at Robert Bosch GmbH, Automotive Electronics, Reutlingen, Germany.

His research interests are centered around microelectromechanical systems.

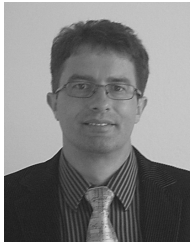

**Jan E. Mehner** received the Diploma and the Dr.-Ing. degree in electrical engineering and information technology from the Chemnitz University of Technology, Chemnitz, Germany, in 1989 and 1994, respectively. From 1998 to 1999 he was a Visiting Scientist at the Massachusetts Institute of Technology, Cambridge, where he was involved in the field of modeling and simulation of MEMS. He is currently Professor for Microsystems and Biomedical Engineering at the Chemnitz University of Technology. His research interests include analytical and numerical methods for microsystems design, design automation, model order reduction and computational algorithms for coupled field analysis as well as model export to different design environments.

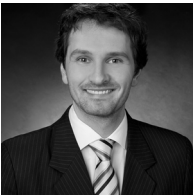

**Peter Degenfeld-Schonburg** received the Diploma and the Dr.-rer. nat. degree in theoretical physics from the Technical University of Munich, Munich, Germany, in 2010 and 2016, respectively. His PhD research was focused on driven and dissipative quantum many-body systems and nonlinear quantum optics. He developed the self-consistent projection operator theory for the simulation of dissipative nonlinear quantum systems.

Currently, he is a research scientist at the Department of Microsystems and Nanotechnology at Robert Bosch GmbH, Corporate Research, Renningen, Germany. His responsibilities include the design and modelling of MEMS sensors and actuators with special emphasis on the simulation of nonlinear dynamical effects.